# A Sweet Recipe for Consolidated Vulnerabilities: Attacking a Live Website by Harnessing a Killer Combination of Vulnerabilities for Greater Harm

Mazharul Islam[1], MD. Nazmuddoha Ansary[2], Novia Nurain[3], Salauddin Parvez Shams[4], A. B. M. Alim Al Islam[5]

[1, 3, 5] Department of Computer Science and Engineering
[2, 4] Department of Electrical and Electronic Engineering
[1, 3] United International University, Dhaka, Bangladesh
[2, 4, 5] Bangladesh University of Engineering and Technology, Dhaka, Bangladesh
Email: [1]mazharul@cse.uiu.ac.bd, [2]nazmuddoha.ansary.28@gmail.com,
[3]novia@cse.uiu.ac.bd, [4]parvezshams38@gmail.com, [5]alim razi@cse.buet.ac.bd

*Abstract*—Recent emergence of new vulnerabilities is an epoch- making problem in the complex world of website security. Most of the websites are failing to keep updating to tackle their websites from these new vulnerabilities leaving without realizing the weakness of the websites. As a result, when cyber-criminals scour such vulnerable old version websites, the scanner will represent a set of vulnerabilities. Once found, these vulnerabilities are then exploited to steal data, distribute malicious content, or inject defacement and spam content into the vulnerable websites. Furthermore, a combination of different vulnerabilities is able to cause more damages than anticipation. Therefore, in this paper, we endeavor to find connections among various vulnerabilities such as cross-site scripting, local file inclusion, remote file inclusion, buffer overflow CSRF, etc. To do so, we develop a Finite State Machine (FSM) attacking model, which analyzes a set of vulnerabilities towards the road to finding connections. We demonstrate the efficacy of our model by applying it to the set of vulnerabilities found on two live websites.

## I. INTRODUCTION

According to Common Vulnerability Explorers (CVE) [2], in the last two years, i.e., 2018 and 2017, the number of vulnerabilities found on websites are more than 11, 000. These massive number of vulnerabilities leave a insidious attacking surface for malicious hackers. Since vulnerabilities are the attacking surface for the hackers, more websites are becoming vulnerable to hacking. Consequently, most website developers give little efforts to maintain the security of website updated. They are not vigilant enough to keep updating the website security regularly.

Furthermore, most of these vulnerable websites have compelling ubiquitous effect in personal, social, and economic life as these websites contain sensitive information from credit card pin number to passwords. If any one of these sensitive information falls into the hand of malicious hacker results can be devastating.

Hence, exploration of vulnerabilities of a website has always been a study of interest to the research community. There are many scanners available such as Acunetix [1], Arachni [3], Netspark [4], etc., to scan live websites for exposing vulnerabilities. Such vulnerabilities includes SQL Injection, Cross Site Scripting, Cross-Site Request Forgery, Local File Inclusion, and Remote File Inclusion. However, Cross Site Scripting and SQL injection have been the main focus of interest owing to more malicious effects.

From the attackers point of view, attackers envision to relay on these vulnerabilities in order to construct malicious attacks so that he can penetrate the security of live websites. However, we explore a potential gap here.When we are only considering an isolated vulnerability one at a time, we may end up doing not so great harm to the security of the website. However, the impacts of conducting a series of attacks combining diversified vulnerabilities exposed on websites is yet to be explore in the literature. Therefore, in this paper, we endeavor to exhibit that we can achieve more malicious results by combining vulnerabilities which would have been impossible to do just by considering one isolated vulnerability. Here, to justify our endeavor, we can draw a real-life analogy of a chess game. In an international level chess game, a grand master may combine a series of brilliant moves according to a genius game plan. When analysts look at these series of brilliant moves they say that each isolate move may not do much harm to the opponent but a series of these moves cause irreparable damage to the chances of the opponent winning the game. Similarly, we can combine vulnerabilities and execute a series of attacks combining these vulnerabilities to cause great harm.

In this paper, we develop a Finite State Machine (FSM) attacking model which will connect and combine the found vulnerabilities. Each vulnerability is associated with some preconditions and postconditions. The attacker can take advantage of any particular vulnerability if and only if the precondition statements are true. The successful exploitation of the vulnerability will leave the postconditions to be true.

Now, the new true conditions will enable the attacker to exploit advantages of other vulnerability, which has these new conditions as its preconditions.

Our proposed FSM model has a starting state as the place where the attacker begin to initiate attack. The starting state has no precondition statements. Then, using various attacks and vulnerabilities the attacker try to reach the goal state. The goal states are successful exploitation of the victim website which is impossible to reach using only one isolated vulnerability.

In this paper, we make the following contributions:

- Our proposed FSM attacking model exhibits that combining different vulnerabilities using preconditions and postconditions enable us to reach great harmful goal states which are not possible to reach using only one isolated vulnerability.
- We deploy our propose FSM attacking model against two different real-life live websites. One is http://testphp.vulnweb.com, which is open for penetration testing. The other one is http://teacher.xxx.xx.xx, which is an official government website.
- Our evaluation reveals that our proposed FSM attacking model provides a sequence of attacks combining different exposed vulnerabilities causing devastating effects on the two compromised websites.

The rest of this paper is structured as follows: Section II will highlight related works, Section III will delineate the construction of Finite State Machine (FSM) and our recursive algorithm to reach accepting/goal states. Section IV will give comprehensive analysis of the test results that we have found on two live website using our FSM model. Section V will conclude the paper giving a brief remark on future work.

## II. RELATED WORK

Detection of various vulnerabilities specially SQL injection and cross site scripting is popular in the literature of cyberspace security. Sonewar et. al. propose an approach for detection of SQL injection and cross site scripting attack [5]. The study in [6] investigate on finding SQL injection and cross site scripting using static analysis tool. Besides, existing studies such as [7]–[10] explore cross site scripting and SQL injection separately in the literature pertinent to penetration of website.

The holistic approach of considering all vulnerabilities has inspired other researches such as [11]. However, their research suffers from one major assumption that we will be able to scan the internal network of the victim website which is not practical. Since hackers wont have the required permission to go inside the firewall of the victim websites network and perform scanning.

In this paper, we have taken a holistic approach. Instead of considering each of the attacks separately, we envision to look at the whole picture of available vulnerabilities and to combine them all for causing greater harm.

## III. DESIGN ARCHITECTURE OF ATTACKING MODEL

We design a Finite State Machine (FSM) to connect the existing vulnerabilities on the victim website. We call this stage preprocessing. We choose Finite State Machine (FSM) since the modeling of the problem is quite complex. According to our thinking, Finite State Machine (FSM) can easily consume these type of complexities while modeling this kind of problems.

We model our Finite State Machine (FSM) in such a way so that accepting states (goal states) represent causing severe harm to the victim website and starting state represents the initiation stage before attacking. Our target is to reach accepting states (goal states) from starting state. To summarize, we divide our model into two stages as shown in Fig. 1

- The first stage is for building the Finite State Machine (FSM). This is a preprocessing stage.
- In the second stage, we run our recursive algorithm to attain our goal of directing towards more harm for the victim website.

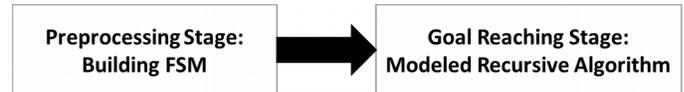

Fig. 1. Two stages of our FSM based attacking model

### A. Preprocessing Stage: Building FSM

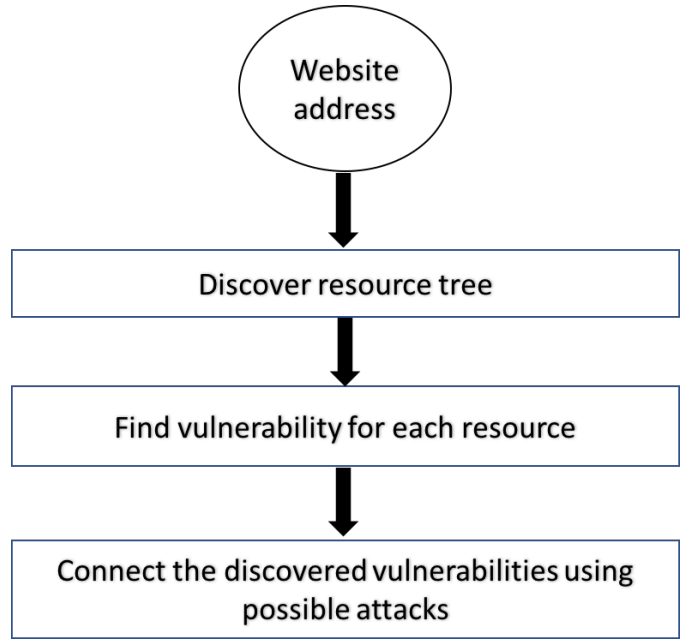

Fig. 2. Three sequential phases of preprocessing stage: building Finite State Machine

We develop our Finite State Machine (FSM) in three sequential phases as shown in figure 2. This preprocessing stage will take as input the victim website address and outputs a Finite State Machine (FSM).

- **Phase 1: Discovering URI tree**: In the first phase, we find every possible accessible resource (URI) to build a knowledge base for the victim website. The knowledge base contains the victim website information such as Apache version, PHP version, open ports, the list of all URI of the victim website, etc.

- **Phase 2: Finding vulnerabilities and possible attacks for each URI:** In the second phase, we use the knowledge base of the first phase to discover vulnerabilities and possible attacks on the victim website. We give the URI list discovered in the first phase to available scanners such as Acunetix [1], Arachni [3], Netspark [4], Nikto [14] etc. The scanners give us a list of possible vulnerabilities and attacks that can be executed on the URI.

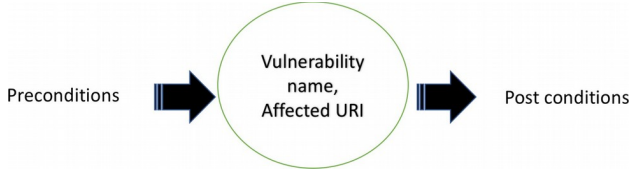

Fig. 3. Four properties of a state: vulnerability name, affected URI, precondition, postcondition

- **Phase 3: Building FSM by dependencies Analysis:** In the third phase, we connect the isolated vulnerabilities using the preconditions and postconditions. We can successfully take advantage of a vulnerability of if all of its preconditions are true. Successful exploitation of the vulnerability will render the postcondition to be true.

We have defined the four properties of our Finite State Machine as follows:

- **State**: To differentiate each state of our Finite State Machine, we have considered each state as a unique combination of a vulnerability and the URI that which is affected by the vulnerability. Each state is associated with some preconditions and postconditions. We define preconditions and postconditions as following:
  - **Preconditions:** The attacks which must be executed successfully to take advantage of the vulnerability of the affected URI.
  - **Preconditions:** The attacks which can be executed on the victim website leveraging the vulnerability of the affected URI.

Thus, each state has four properties as illustrated in the Fig. 3 and in Table I
- **Edges**: Edges in our FSM connects the states. The incoming edges to a state represent preconditions and the outgoing edges from a state represent postconditions.
- **Starting state**: We use the NULL situation before initiate attacking as the starting state. The initial state requires no precondition to be true (starting state has no preconditions). We draw outgoing edges from the starting state to those states which do not need any precondition to be true.
- **Accepting state**: Accepting (goal) states represent a harmful state of the victim website after combining more than one states.

TABLE I
THE FOUR PROPERTIES OF A STATE

| Vulnerability name |
| --- |
| Affected URI |
| Preconditions |
| Postconditions |

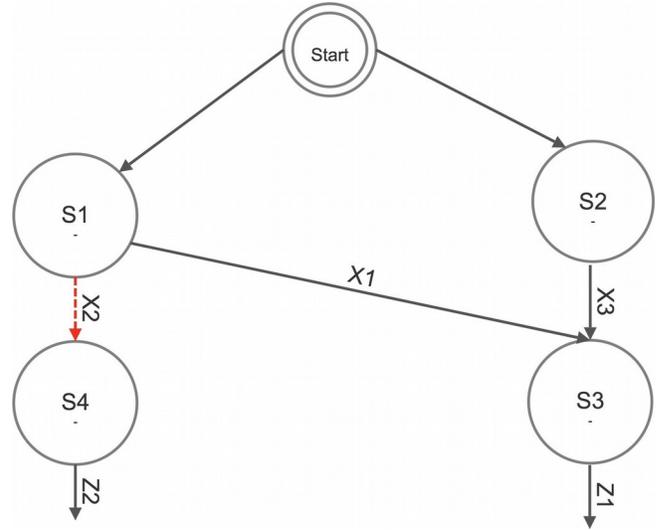

Fig. 4. **Simplified state diagram**

After the preprocessing step, we build a Finite State Machine (FSM) for the target victim website. We summarize the algorithmic details in **Algorithm 1**.

B. Goal Reaching Stage: Modeled Recursive Algorithm

We first mark all states as not visited. Then, we mark starting state as visited and try to visit all other states in using a $DFS(starting\_state)$ function call in a recursive fashion. We can reach a state if all the preconditions along the path to that state are true. After the successful exploitation, the new true postcondition statement will become true. This may enable us to reach those states which were not possible for us to reach because of its precondition being false previously.

After this recursive algorithm, the visited marked goal states signify the harmful damage we can do to the security of the website. We present the process of goal reaching sate in Algorithm 2.

A simplified version of our FSM is shown in Fig. 4. The preconditions and postconditions of the corresponding FSM along with the state are shown in Table II.

We can observe that vulnerability A affects two URI namely a1 and a2. Vulnerability B and C affect URI b1 and c1 respectively. So we need 2 + 1 + 1 = 4 states to identify

**Algorithm 1** Algorithm for Building FSM: Preprocessing Stage

**Input:** victim website address

**Output:** A FSM

  *Initialization* :
1: TrueConditionList ← ∅
2: URITree

3: URIVulnerabilityMap ← ∅
4: **for each** new_resource found in Phase 1 **do**
5:     add the new_resource to URITree
6:     URIVulnerabilityMap [new_resource ] ← ∅
7: **end for**
8: **for each** leafURI in URITree **do**
9:     **for each** vulnerability found in leafURI **do**
10:        URIVulnerabilityMap [leafURI].add(vulnerability)
11:    **end for**
12: **end for**
13: **for each** new_true_condition found in Phase 1 **do**
14:    TrueConditionList.add(new_true_condition)
15: **end for**
16: $FSM$ ← ∅
17: **for each** URI in URITree **do**
18:    **for each** vulnerability in URIVulnerabilityMap[leaf] **do**
19:        preconditions ← ∅
20:        postconditions ← ∅
21:        **for each** condition in TrueConditionList **do**
22:            **if** condition is a prerequisite for vulnerabiltity **then**
23:                preconditions ← preconditions + condition
24:            **end if**
25:            **if** condition is a consequence of vulnerabiltity **then**
26:                postconditions ← postconditions + condition
27:            **end if**
28:        **end for**
29:        State = new State(leafURI, vulnerability, precondition, postcondition)
30:        $FSM$.add($State$)
31:    **end for**
32: **end for**
       initialPostConditions ← ∅
33: **for each** condition not prerequisite for any vulnerabiltity **do**
34:    initialPostConditions ← initialPostConditions + condition
35: **end for**
36: $Start$ = new $State$ (rootURI, NULL, NULL,)
37: $FSM$.add($Start$)

---

**Algorithm 2** Algorithm for Goal Reaching: Recursive Stage
**Input:** The $FSM$ built in the preprocessing stage
**Output:** A set of goal states reachable from starting stage
   *Initialization*:
1: all_states ← Not visited
2: starting_state ← visited
3: $DFS(starting\_state)$
4: $goal\_set$ ← ∅
5: **for each** $state$ in FSM **do**
6:    **if** $state$ == Visited **then**
7:        $goal\_set$. add($state$)
8:    **end if**
9: **end for**
10: **return** $goal\_set$
    $DFS(State)$
11: Mark all postconditions of $state$ True
12: **for each** $state$ in FSM **do**
13:    **if** $state$ == Not visited **then**
14:        **if** All of its precondition are true **then**
15:            $state$ ← visited
16:            **for each** $condition$ in post-condition of $state$ **do**
17:                $condition$ ← True
18:            $DFS(state)$
19:            **end for**
20:        **end if**
21:    **end if**
22: **end for**

TABLE II
VULNERABILITY PRECONDITIONS AND POSTCONDITIONS

| Sate | Vulnerability | URI | Preconditions | postconditions |
|------|---------------|-----|---------------|----------------|
| S1   | A             | a1  | -             | x1, x2         |
| S2   | A             | a2  | -             | x3             |
| S3   | B             | b1  | x1, x3        | z1             |
| S4   | C             | c1  | x2            | z2             |

each vulnerability and URI pair uniquely. The starting state S0 has no precondition. We connect the starting state with those states which do not need any preconditions to be true which in this case from Table II are S1 and S2.

We can conclude from Fig.4 that to reach S3 we need the preconditions conditions x1 and x3 to be true. x1 is the postcondition of S1 and x3 is the postcondition of S2. So to reach S3 we need to take advantage of vulnerability A on URI a1 and a2. If we can make the postcondition of S1 and S3 true by successfully executing attacks on URI a1 and a2 by leveraging vulnerability A, only then we can reach S3.

We should make a remark at this point. The possible vulnerabilities and attack list to make postconditions true by leveraging some vulnerability on a URI are given to us by the scanners. Since scanners are not 100% accurate, there may exist false positive attacks. Such as we can see in figure 4 that in spite of reaching **A**, we still can not make postcondition **x2** true because the possible attacks to make this postcondition x2 true are false positive attack reported by the scanners. The dotted red arrow **x2** represents false positive attacks in Fig. 4

## IV. EXPERIMENTAL RESULTS

We choose two different real-life live websites http://testphp.vulnweb.com.com (open for penetration testing)

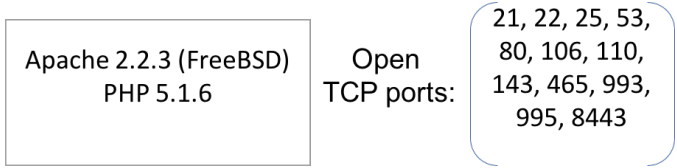

Fig. 5. Outcomes of executing phase 1 on http://testphp.vulnweb.com in terms of server information

and http://teacher.xxx.xx.xx (public government website hence we are being anonymous about the website address) for testing our FSM attacking model. In both cases, our attacking machine is Kali linux (2018 July release version). We download the virtual machine images of Kali linux and run these image files using Oracle Virtual Box.

### A. Results on http://testphp.vulnweb.com

Acunetix developed http://testphp.vulnweb.com for testing purposes. It is a vulnerable and dummy website authorized for penetration testing only. The website is fairly simple functionality wise. It contains a list of dummy artists' name and there are some dummy artworks with the artist's name. After logging in the user can buy and add these artworks to his cart.

In order to compromise the security of the website http://testphp.vulnweb.com our FSM attacking model leverages the vulnerabilities found on the website and relates these vulnerabilities in two stages as shown in figure 1.

We will first discuss the three sequential phases of the preprocessing stage to build the FSM.

**Outcomes of phase-1**: In this phase, we gain as much information as we can about the website. We are able to crawl the entire website resources in this case. For crawling and fetching HTML, PHP and XML data of the website, we utilize the Python framework BeautifulSoup [12]. We use Nmap [15] for scanning open ports. Figure[5] shows the server information and figure[6] shows the resource tree of our target victim website http://testphp.vulnweb.com.

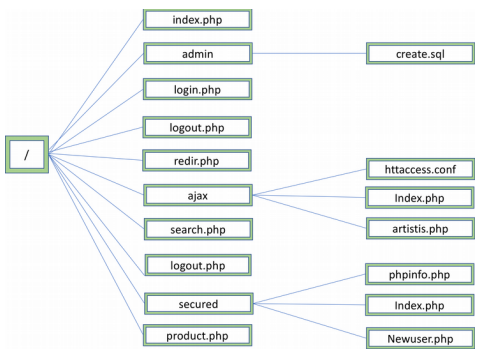

Fig. 6. Outcomes of phase-1 on http://testphp.vulnweb.com in terms of URI resources. The leaf nodes represents resources. The non-leaf nodes represents directory/folder

**Outcomes of phase-2**: Outcomes of phase-2: After having, the knowledge base of the website at our disposal, we then find the vulnerabilities of each URI as shown in figure[6] in the second phase.

We give resources of the website as an input to scanners. The scanners find their vulnerabilities along with the necessary preconditions and postconditions. The results are of the scan is summarized in figure III.

**Outcomes of phase-3**: This phase is the most crucial part of our model. In this phase, we build the Finite State Machine (FSM). We build the FSM by defining the four properties of an FSM:

- **Sates**: As shown in Table IV we denoted 10 states each being a unique combination a vulnerability and a URI.
- **Edges**: We have associated each state with some precondition and post-condition. The preconditions and postconditions are shown in Table III are used to connect the 10 states. The number of incoming edges to a state and outgoing edges from a state is respectively equal to the number of preconditions and postconditions the corresponding state has. Such as, in Fig.7, S6 has 3 preconditions and 1 postcondition. Hence, we associate 3 incoming edges and 1 outgoing edge with S6.
- **Starting state**: S0 represents the starting state where we begin to initiate series of attacks. We connect S0 with those states which have no preconditions. Such as Fig. 7 shows that S0 is connected to S1, S2, S3, S5, S9 as these states have no preconditions in them.
- **Accepting state**: The RED colored states represents a goal/accepting state of the FSM. Such as Fig. 7 shows S4, S7, S10 are the goal states here. S4, S7, S10 are the goal states not reachable from S0, if we had not considered other states and connected these states using preconditions and postconditions.

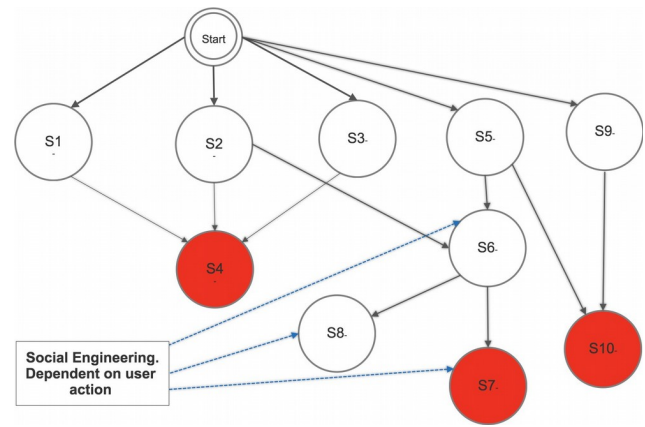

Fig. 7. The simplified Finite State Machine (FSM) for http://testphp.vulnweb.com The states are connected using the preconditions and postconditions from Table IV

For the demonstrative purpose, we are going to delineate how we can reach S4 from S0.

TABLE III
OUTCOMES OF PHASE-3 ON http://testphp.vulnweb.com IN TERMS OF BUILDING A FSM

| State | Vulnerability | URI | Precondition | Postcondition |
|---|---|---|---|---|
| S1 | Weak password | /login.php | None | Narrow search space of password. |
| S2 | Email addresses discloser | /index.php | None | Email of a registered is found. |
| S3 | HTTP basic authentication | /auth.php | None | Not blocking multiple failed log-in attempts. Not blocking the same GET request again and again. |
| S4 | Brute force dictionary attacks possible | /auth.php | Narrow search space of password. Email address of a registered user Not blocking multiple failed login attempts | Successfully logging in as registered user. |
| S5 | iFrame header is missing | /login.php | None | Embedding the logging form into a third party web-page. |
| S6 | Social engineering attacks | /login.php | Embedding the logging form into a third party web-page User clicking the address of third party web-page sent in email Defining emails of registered users | User redirected to a third party web-page |
| S7 | Cross-site Request Forgery in Login Form | /login.php | Redirection to a third party web-page User filling up the logging form | Reading the user name password entered by the user from the third party |
| S8 | Session-cookie without HttpOnly | ALL URI | Redirection to a third party web-page Session cookie already save in client browser for the logged in user | Stealing session cookie PHPSESSID of logged in user from thrid part web-page |
| S9 | Slow response time | /Flash/add.fla | None | Running time and memory expensive process for the server |
| S10 | DDoS | /Flash/add.fla | Running time and memory expensive process for server Not blocking any request to load the same resource again and again | END |

**From S4 to S0**: To reach S4 from So, as we can from Fig. 7 we need 3 preconditions to be true. From Table III we can see that these three conditions are.

- **Weak Password**: The scanner reports that password set by the registered users have weak passwords. This is a good news for us since it increases our chances of cracking the password by trying all passwords from dictionary password lists. The outgoing edges of S1 make this precondition true for S4.
- **Email address discloser**: The scanner found a registered users email address from 17 pages in total. This email address is useful since now we don't need the usernames of the registered users. This website allows the registered users to log into the website using emails.
- **Not blocking multiple login attempts**: This website uses HTTP basic authentication which is vulnerable since it does not block users for multiple failed login attempts. Our brute force Python script tries logging in one attempt per second. Still, the server did not block us, after multiple failed login attempts because one of the postcondition of S3 is not to block.

Since all these preconditions are true, now we can conclude that the victim website is vulnerable to brute force dictionary attacks affecting /auth.php URI meaning we can reach state S4.

In the same way, we can say the same things about other states. Fig. 7 is drawn using the Table III. Interesting to note as shown in Fig. 7, S6, S7, and S8 have a dotted incoming edge. These edges mean that the value of these preconditions information and Fig. 9 shows the resource tree of the victim live website http://teacher.xxx.xx.xx.

to be true, we have to rely on victim user fault or carelessness. Such as the dotted incoming edge to S6 represents if the user has clicked the malicious link we have sent to him. If the user clicks, then this precondition becomes true. A careful user would not click the link in the email making this precondition false.

### B. Results on http://teacher.xxx.xx.xx

The reason for choosing http://teacher.xxx.xx.xx to test our FSM attacking model is that it is an outdated website. However, being a government website, it is used for such important official work. Executing active attacks against this official website requires higher order permission which was not possible for us to get before the paper submission deadline. Hence we have only executed passive attacks.

Therefore we have used our Finite State Machine (FSM) attacking model in passive mode only. We only executed those attacks which would not cause any direct harm to the website. In the preprocessing stage, we build an FSM in three sequential steps as shown in figure 2. In the goal reaching stage, we reach those states which were not possible to reach considering only isolated states. We are going to describe the outcomes of the three stages of prepossessing, then present the goal reaching stage.

**Outcomes of phase-1**: For gaining as much information and resources as we can about the website. we crawl the entire victim website. For crawling and fetching the resources of the website, we utilize Python framework BeautifulSoup [12]. We use Nmap [15] for scanning open ports. Fig. 8 shows server

Fig. 8. Outcomes of phase-1 on http://teacher.xxx.xx.xx in terms of server information.

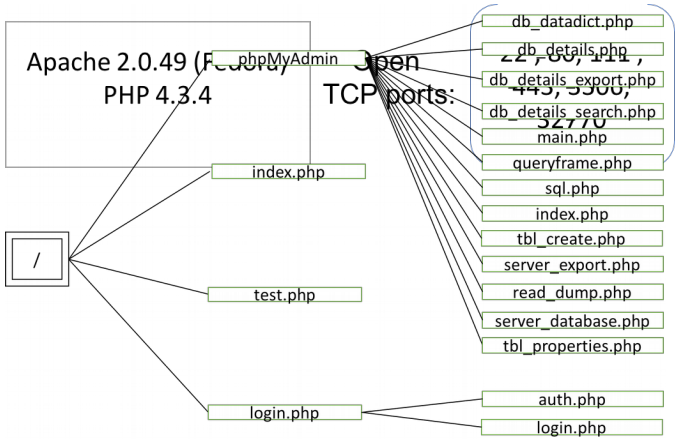

Fig. 9. Outcomes of phase-1 on http://teacher.xxx.xx.xx in terms of URI resources. The leaf nodes represents resources. The non leaf nodes represents directory/folder

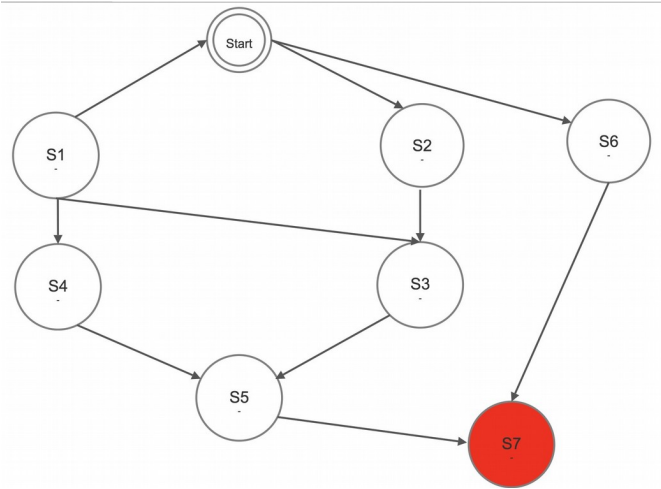

Fig. 10. Outcomes of phase-3: build FSM on http://teacher.xxx.xx.xx

**Outcomes of phase-2**: After gaining the knowledge base of the website, we then find the existing vulnerabilities for URI resources on the outdated live website http://teacher.xxx.xx.xx in this second phase.

In this phase, we utilize Acunetix [1], Netspark [4], Nikto [14], and Dirbuster [13]. We use four scanners for double checking of the existing vulnerabilities. Found vulnerabilities after double checking are summarized in Table IV.

**Outcomes of phase-3**: This the phase where we connect vulnerabilities and build the FSM. As we can observe from Table IV, We denote 7 states. Each state is a unique combination of a vulnerability and URI resource. The build FSM from the resulted table IV of phase-3 is shown in Fig 9.

Now the preprocessing is complete and we have an FSM on our hand, we are going to explain how can we reach goal state S7 using our recursive algorithm as shown in Algorithm 1.

**From Start to S3**: S3 represents the state of /phpMyAdmin/index.php page being vulnerable to insufficient sanitation of user-supplied input. We can see it has two incoming edges representing two preconditions namely MySQL being the user database and PHP version $\leq 2.x.x$. Both of the preconditions are true. Post-condition of S1 ensures MySQL is the user database. Postcondition of S2 reveals that PHP version is 2.0.49 ($2.0.49 \leq 2.x.x$). Hence we can reach S3 from Start since S1 and S2 has no preconditions.

**From S3 to S5**: When we reach S3, we mark its postconditions as true. Marking the only post-condition of S3 means we can use the set variable such as $what$ equal to the file we want to see. However to view the file we need to log in as a user. We can only see those files which have read permission for the logged in user. To log in as non-privilege user we can advantage of the post-condition of S4. We can mark S4 as visited since we have already visited S1 before while exploring S3. As a proof of concept, we are attaching the screen shot of the contents of */etc/passwd* file which is the post-condition of S5.

**From S5 to S7**: S7 represents a state where we can log in as a registered user. We can visit S7 is the two preconditions are true. Looking at the precondition of S7 we see one of them getting usernames of registered users which is true since we can visit S5. Also, We can execute a python script to brute the password since the HTTP basic authentication is used and it doesn't prevent us from trying to log in to an account with the same, again and again, meaning a successful brute force dictionary attack. Now, this brute force is an active attack and if we want to execute a brute force attack we need permission which we do not have. Hence we could not simulate in real life if we can actually reach S7.

We are presenting the proof of concept of the two preconditions of S7.

- The first precondition possible brute force attack is a postcondition of the S5 stage which is possible brute force attack. Since HTTP basic authentication is used we suspect a possible attacking surface for brute force attack. Indeed when we are run our brute force Python script is was true as shown in Fig. 11. We execute a brute force attack against http://teacher.xxx.xx.xx in less aggressive mode since we were not allowed to harm the website in any way. We only need the proof of concept of a possible brute force attack.

TABLE IV
OUTCOMES OF PHASE-3 ON HTTP://TEACHER.XXX.XX.XX IN TERMS OF BUILDING A FSM

| State | Vulnerability name | URI | Precondition | Post-condition |
|---|---|---|---|---|
| S1 | PHPinfo() page found | /test.php | None | Administrative URI of DB is /php-MyAdmin/index.php. Deafult user account and password has no value |
| S2 | Out-of-date Version (Apache) | NULL | None | PHP version 2.0.49 |
| S3 | Insufficient sanitation of user-supplied input | /phpMyAdmin/index.php/ | MySQL must be used as DB. PHP version ≤ 2.x.x | revealing the contents of directories . to remote attackers |
| S4 | Unauthorized logging | /phpMyAdmin/index.php | Deafult user account and password has no value | Successful logging in as non privileged user. |
| S5 | Local file inclusion | /phpMyAdmin/export.php?what=../../../../../../../../../../etc/passwd%00 | revealing the contents of directories to remote attackers. Successful login as non privileged user | Getting user names of registered users by accessing file /etc/passwd/ which have permission 777. |
| S6 | HTTP basic authentication | login/auth.php | | Possible brute force attack |
| S7 | Unauthorized logging | /login/auth.php | Getting usernames of registered users. HTTP basic authentication | Successfully log in as registered users. |

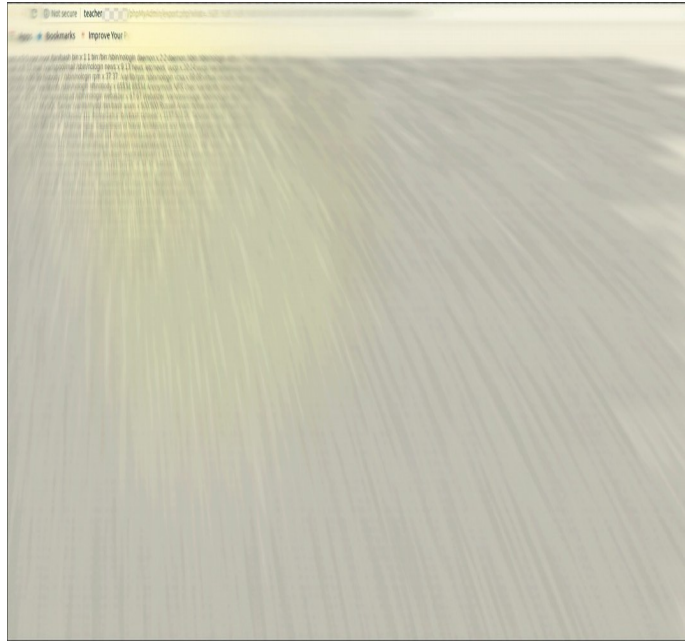

Fig. 11. Proof of concept: possible brute force attack (in less aggressive mode) on http://teacher.xxx.xx.xx

Fig. 12. Proof of concept: getting user names of registered user by accessing file /etc/passwd on http://teacher.xxx.xx.xx .The picture is blurred since it contains sensitive information such as registered usernames and their home directory location.

- The other precondition is to get the usernames of registered users. Fig.12 show the contents of /etc/passwd which is a post-condition of the state S5. The picture is blurred since it contains sensitive information such as registered usernames and their home directory location.

## V. FUTURE WORK

In the future, we have ambitions to automate the entire process of connecting the vulnerabilities and finding out ways to cause more harms to victim websites. Consequently, we want to make our FSM model dynamic so that it can adapt itself with the introduction of new types of vulnerabilities together. Deployment of our FSM attacking model on two live websites illustrates the effectiveness of our proposed model by discovering more harmful states. Consequently, our proposed leveraging techniques of deep learning. Moreover, we will add new metrics such as time, amount of sensitive information obtained, etc., to measure how much damage our FSM model can do to a website.

## VI. CONCLUSIONS

In this paper, we endeavor to find the holistic approach to connect all possible vulnerabilities in order to cause more harm to the security of the websites in comparison to harm caused by one or two vulnerability. Therefore, in this paper, we propose an FSM attacking model combine different vulnerabilities

FSM attacking model acts as a security assessment tool for any websites. Moreover, our FSM attacking model is one of the steps towards the direction of automating the website security assessment tools.

## References

[1] Acunetix Web Scanner, https://www.acunetix.com/, Last Accessed 23 12 2011.
[2] common vulnerability explorers, http://cve.mitre.org/, Last Accessed 8 9 2018.
[3] Arachni Web Application Security Scanner Framework http://www.arachni-scanner.com/, Last Accessed 8 9 2018.
[4] NetSpark — Technology for a Safer Internet http://netspark.com/, Last Accessed 8 9 2018.
[5] Sonewar, Piyush A., and Nalini A. Mhetre. "A novel approach for detection of SQL injection and cross site scripting attacks." In Pervasive Computing (ICPC), 2015 International Conference on, pp. 1-4. IEEE, 2015.
[6] Algaith, A., P. Nunes, J. Fonseca, I. Gashi, and M. Viera. "Finding SQL Injection and Cross Site Scripting Vulnerabilities with Diverse Static Analysis Tools." (2018).
[7] Ismail, O., Etoh, M., Kadobayashi, Y., & Yamaguchi, S. (2004). A proposal and implementation of automatic detection/collection system for cross-site scripting vulnerability. In Advanced Information Networking and Applications, 2004. AINA 2004. 18th International Conference on (Vol. 1, pp. 145-151). IEEE.
[8] Gupta, S. and Gupta, B.B., 2017. Cross-Site Scripting (XSS) attacks and defense mechanisms: classification and state-of-the-art. International Journal of System Assurance Engineering and Management, 8(1), pp.512-530.
[9] Som, S., Sinha, S. and Kataria, R., 2016. Study on sql injection attacks: Mode detection and prevention. International Journal of Engineering Applied Sciences and Technology, Indexed in Google Scholar, ISI etc pp.23-29.
[10] Nagpal, Bharti, Nanhay Singh, Naresh Chauhan, and Angel Panesar. "Tool based implementation of SQL injection for penetration testing." In Computing, Communication and Automation (ICCCA), 2015 International Conference on, pp. 746-749. IEEE, 2015.
[11] Jajodia, Sushil, Steven Noel, and Brian OBerry. Topological analysis of network attack vulnerability. In Managing Cyber Threats, pp. 247266. Springer, Boston, MA, 2005.
[12] Beautiful Soup Documentation https://www.crummy.com/software/BeautifulSoup/bs4/doc/, Last Accessed ,Last Accessed 8 9 2018.
[13] Dirbuster(URLfuzzer:OWASP) https://www.owasp.org/index.php/ Category:OWASP DirBuster Project, Last Accessed 8 9 2018.
[14] Nikto web scanner https://cirt.net/Nikto2, Last Accessed 8 9 2018.
[15] Nmap: the Network Mapper- Free Security Scanner, https://www.nmap.org/, Last Accessed 8 9 2018.